\documentclass[aps,prx,bibliography,superscriptaddress,,twocolumn]{revtex4-2}
\usepackage[utf8]{inputenc}
\usepackage{amsmath}
\usepackage{physics}
\usepackage{mathtools}
\usepackage{amsfonts}
\usepackage{amssymb}
\usepackage{braket}
\usepackage{graphicx}
\usepackage{subfigure}
\usepackage{textcomp}
\usepackage{color}
\usepackage[dvipsnames]{xcolor}
\usepackage{mathrsfs}
\usepackage{natbib}
\usepackage{makecell}
\usepackage{soul}

\begin{document}
\title{Hexagonal Warping Control of Exceptional Points in Topological Insulator–Ferromagnet Heterojunctions}
\author{Md Afsar Reja}
\email{afsarmd@iisc.ac.in}
\affiliation{Solid State and Structural Chemistry Unit, Indian Institute of Science, Bangalore 560012, India}
\author{Awadhesh Narayan}
\email{awadhesh@iisc.ac.in}
\affiliation{Solid State and Structural Chemistry Unit, Indian Institute of Science, Bangalore 560012, India}
\date{\today}
\begin{abstract}
Exceptional points (EPs) are non-Hermitian degeneracies, where both eigenvalues and eigenvectors coalesce, which are fundamentally distinct from their Hermitian counterparts. In this study, we investigate the influence of hexagonal warping on EPs emerging at the interfaces between topological insulators and ferromagnets. We demonstrate that the presence of the warping term plays a crucial role in determining the locations of the EPs. Furthermore, we show that the number as well as the positions of EPs emerging at such junctions can be tuned by an applied magnetic field. Our results, in line with previous studies on topological insulator-ferromagnet junctions, suggest them as a promising platform for realizing non-Hermitian physics.

\end{abstract}

\maketitle

\section{Introduction}

\textcolor{black}{The emergence of Topological insulators (TIs) has driven enormous research activity in the past two decades. TIs host a plethora of fascinating properties, such as protected edge and surface states, spin-momentum locking, unconventional transport properties, to name just a few~\cite{hasan2010colloquium,qi2011topological}. In addition, interfacing them with different quantum phases can result in intriguing phenomena. For instance, interfacing TIs with superconductors is expected to realize Majorana quasiparticles~\cite{lee2018topological,alspaugh2021majorana}, while their proximity with magnets can lead to half-integer quantum Hall effects~\cite{wu2009half,konig2014half}.} One of the intriguing properties of topological insulators is the emergence of snowflake-shaped Fermi surface states, in contrast to the usual circular Fermi surface~\cite{chen2009experimental,hsieh2009tunable}. The underlying crystal symmetry gives rise to this distinctive Fermi surface geometry, which can be captured by a cubic-in-momentum $k^3$ hexagonal warping term~\cite{fu2009hexagonal,basak2011spin,alpichshev2010stm}. The role of hexagonal warping effects has previously been explored in various contexts, including optical conductivity~\cite{li2013hexagonal}, circular dichroism~\cite{li2014hexagonal}, nonlinear Hall effects~\cite{saha2023nonlinear}, multi-Weyl semimetals~\cite{chowdhury2022exceptional}, and transport properties~\cite{wang2011effects,choudhari2019effect}.\\

\textcolor{black}{In recent years, topology in the non-Hermitian (NH) regime has received significant attention.} Along with extended classes of topological phases
NH systems exhibit a broad range of novel phenomena, including EPs, the non-Hermitian skin effect, to name a few~\cite{bender2007making,ashida2020non,bergholtz2021exceptional,kawabata2019symmetry,banerjee2023non,ding2022non,el2018non,okuma2023non,meng2024exceptional}. A key feature of non-Hermitian systems is the presence of EPs -- degeneracies at which both eigenvalues and eigenvectors coalesce together~\cite{heiss2012physics,kato2013perturbation}. These are distinct from Hermitian degeneracies, where only the eigenvalues coincide while the eigenvectors remain orthogonal. EPs are connected to a wide range of intriguing phenomena, spanning both theory and practical applications. These include NH topological phases characterized by the winding of eigenvalues and eigenvectors around the EPs~\cite{ding2022non,banerjee2023tropical,jaiswal2023characterizing}, potentially enhanced sensing capabilities~\cite{hodaei2017enhanced}, exceptional optical microcavities~\cite{chen2017exceptional}, and directional lasing technologies~\cite{peng2016chiral}. Experimentally, EPs have been realized in a variety of platforms, including optics, photonics, electrical circuits, and acoustic systems~\cite{parto2020non,miri2019exceptional,stehmann2004observation,choi2018observation,zhu2018simultaneous,shi2016accessing,zhu2018simultaneous,xiao2021observation}. \textcolor{black}{
Several classes of exceptional degeneracies have been studied in heterojunctions composed of superconductors~\cite{pikulin2012topological,pikulin2013two,cayao2023bulk,cayao2024non,cayao2024Phase,li2024anomalous,beenakker2024josephson}. Furthermore, the role of non-Hermiticity has been investigated in the context of Hall transport phenomena~\cite{ling2016anomalous,chen2018hall,philip2018loss}.} Recently, appearance of EPs and their extended versions has been investigated theoretically in heterojunctions composed of unconventional quantum materials and ferromagnets~\cite{kozii2024non,bergholtz2019non,cayao2023exceptional,reja2024emergence,dash2025fingerprint,reja2025n,alipourzadeh2025opto}.

\textcolor{black}{Exceptional degeneracies have been investigated in heterojunctions consisting of ferromagnetic leads and a TI described by a linear-in-momentum Hamiltonian in Refs.~\cite{bergholtz2019non,bergholtz2021exceptional}. In the present work, we consider a more general model of a TI by including the hexagonal warping term to study the fate of the non-Hermitian degeneracies. Specifically, we investigate the role of hexagonal warping in the emergence of NH phenomena at a TI–ferromagnet (FM) junction. We find that the hexagonal warping leads to unique effects on the EPs. Specifically, the warping term splits the exceptional ring into six distinct EPs. The positions of the EPs directly reflect the symmetry imposed by the hexagonal warping. Furthermore, we propose and demonstrate that the EP locations can be tuned by an applied magnetic field and can even be annihilated through field control. We complement our analytical results for the junctions by means of numerical characterization of the NH degeneracies.} Our study reinforces previous studies that TI–FM heterostructures offer a versatile and experimentally accessible platform for investigating non-Hermitian physics. \\

\begin{figure}[t]
\centering
\includegraphics[width=0.32\textwidth]{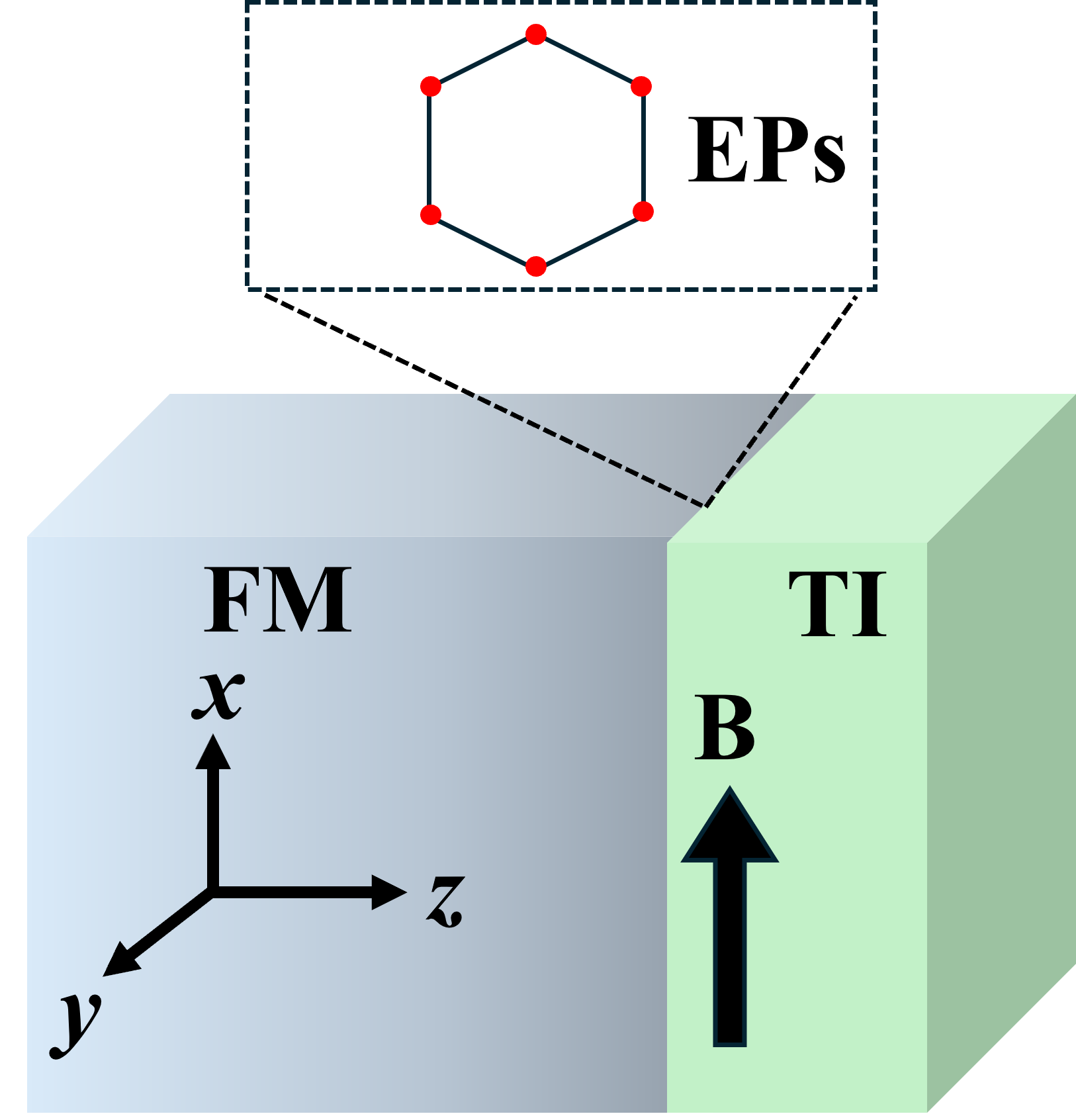}
\caption{\textbf{Schematic of the proposed topological insulator-ferromagnet junction.} The junction is formed at $z=0$, while the region $z<0$ represents the FM lead which is in contact with a TI film (extending for $z>0$). The red dots in the inset denote the emergent EPs, which follow the symmetry imposed by the hexagonal warping term. The thick black arrow denotes the direction of the externally applied magnetic field, which serves as a tuning parameter for controlling the position and evolution of the EPs.}
\label{schematic}
\end{figure}

\begin{figure}[b]
\includegraphics[width=0.45\textwidth]{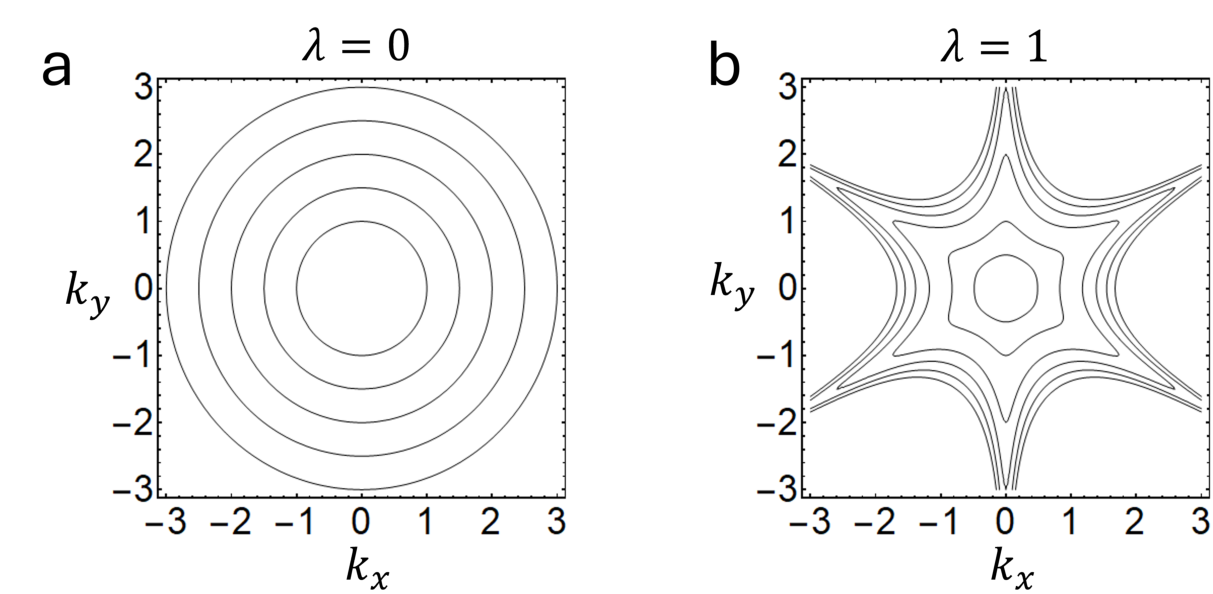}
\caption{\textbf{Comparison of Fermi surfaces with and without the warping term.} 
(a) Circular Fermi surfaces in the absence of the warping term ($\lambda = 0$). 
(b) The Fermi energy contours acquire a snowflake-like hexagonal shape in the presence of the warping term ($\lambda = 1$). The different contours indicate different Fermi energies. Here, we choose $\alpha = 1$.}
\label{Fermi_surface}
\end{figure}

\section{Setup of topological insulator-ferromagnet junction}

We consider a TI coupled to a semi-infinite FM lead, as illustrated in Fig.~\ref{schematic}. The interface is located at $z=0$, with the FM lead occupying the region $z<0$. The resulting junction is treated as an open quantum system and is described by the following effective Hamiltonian,

\begin{equation}
\Tilde{H} = H + \Sigma_L.
\label{eq_H_NH}
\end{equation}

Here, $H$ denotes the Hamiltonian of the TI (introduced below), while $\Sigma_L$ is the self-energy induced by the semi-infinite FM lead. Within the wide-band approximation, the self-energy is independent of both momentum and frequency and can be evaluated analytically~\cite{ryndyk2009green,datta1997electronic,bergholtz2019non,cayao2022exceptional}. \textcolor{black}{Following the approach of Ref.~\cite{cayao2022exceptional} by Cayao \textit{et al.}, we write the self-energy term as }  ,

\begin{equation}
\Sigma_L=-i\Gamma\sigma_0 -i\gamma\sigma_z,
\label{Eq_self_energy}
\end{equation}

where $\Gamma = \frac{\Gamma_{+} + \Gamma_{-}}{2}$, $\gamma = \frac{\Gamma_{+} - \Gamma_{-}}{2}$, and $\Gamma_{\pm} = \pi |t'|^2\rho_{\pm}^L$. Here $\rho_{\pm}^L = \frac{1}{\textcolor{black}{t_z}\pi} \sqrt{1 - (\frac{\mu_L \pm m}{2t_z})^2}$. The quantities $\rho_{\pm}^L$ represent the surface density of states of the lead for the spin-up and spin-down channels, respectively. The parameter $t'$ represents the hopping amplitude between the FM lead and the TI. Here, $\sigma_x$, $\sigma_y$, and $\sigma_z$ are the Pauli matrices, and $\sigma_0$ represents the $2 \times 2$ identity matrix. \textcolor{black}{Pauli matrices are associated with the spin degree of freedom.} 
The parameter $t_z$ corresponds to the hopping amplitude along the $z$-direction within the lead. The quantity $\mu_L$ is the chemical potential of the lead, while $m$ characterizes the intrinsic magnetization of the FM lead. The coupling to the lead makes the effective Hamiltonian of the junction display an NH character through the imaginary component of the self-energy. \textcolor{black}{Coupling to any substrate will cause non-Hermitian effects, such as broadening of energy levels.
The unique aspect of the coupling to the substrate examined in this work is that the coupling
differs for the two spin species. This leads to the emergence of exceptional points and topological
non-Hermitian effects}. We next investigate the exceptional physics at this proposed junction. \\

\begin{figure*}[t]
\centering
\includegraphics[width=0.80\textwidth]{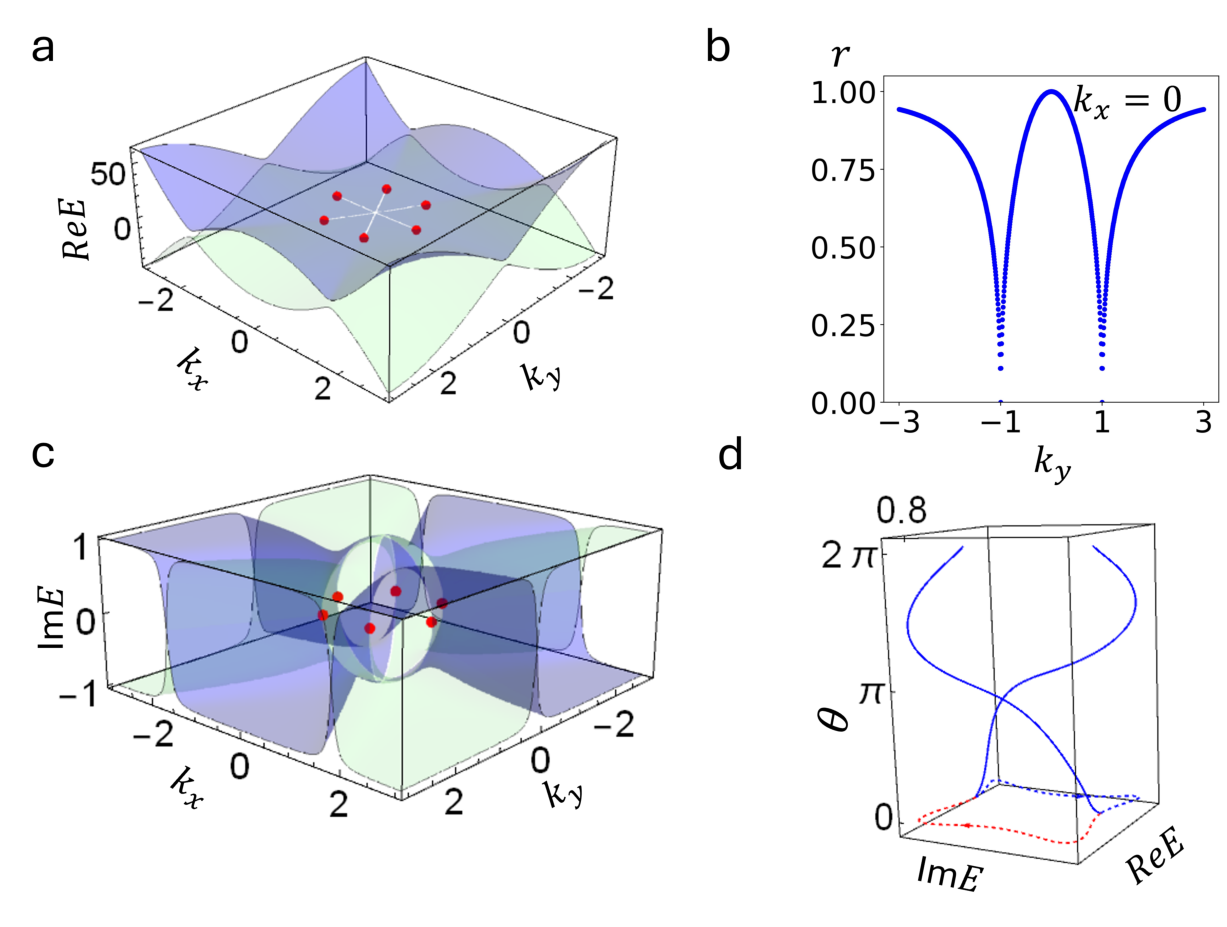}
\caption{\textbf{Complex band diagram of the junction.} The (a) real and (b) imaginary parts of the energy for the TI-FM junction. The two bands are presented in green and blue colors. Note that both the real and the imaginary parts of the eigenvalues coalesce at the red points, indicating the presence of six EPs which exhibit a hexagonal symmetry.(c) Phase rigidity along the $k_x = 0$ line. The vanishing of $r$ as $(k_x,k_y) \to (0,\pm 1)$ confirms the presence of EPs. (d) Vorticity around the EP at $(0,-1)$. The winding of the eigenvalues in the complex energy plane confirms the existence of the EP.    Here, we choose $\lambda=1$, $\alpha=1$,$\gamma=1$, $B_x=0$.}
\label{evals}
\end{figure*}

\begin{figure*}[t]
\includegraphics[width=0.80\textwidth]{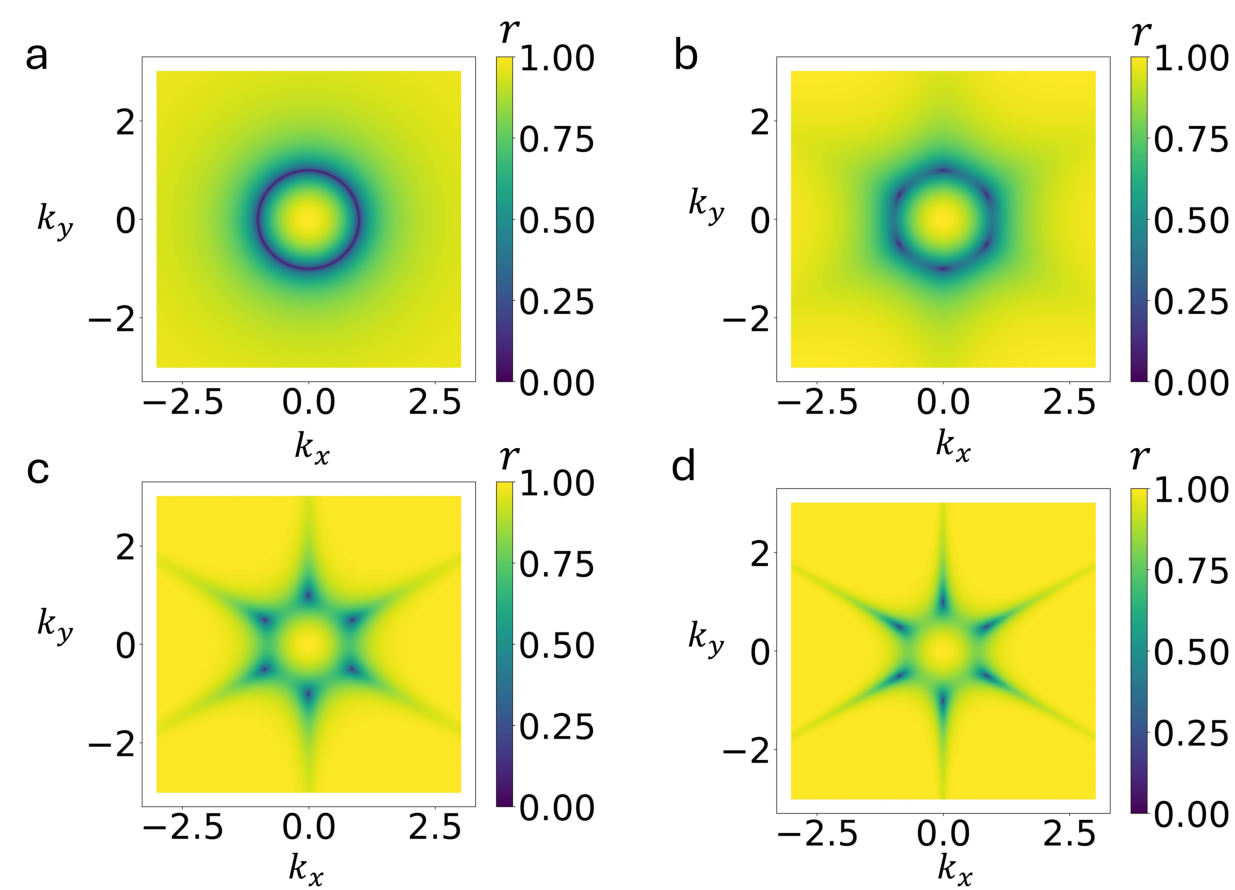}
\caption{\textbf{Role of hexagonal warping in shaping exceptional degeneracies.} The phase rigidity in the $k_x-k_y$ plane with (a) $\lambda = 0.0$, (b) $\lambda= 0.1$, (c) $\lambda = 0.5$, and (d) $\lambda = 1.0$. At zero hexagonal warping, an exceptional ring is formed centered at the origin. Increasing $\lambda$ drives the splitting of the exceptional ring into discrete EPs, reflecting the underlying hexagonal symmetry of the warping term. The EPs are robust to the strength of the hexagonal warping. Here, we choose  $\alpha=1$,$\gamma=1$, $B_x=0$.}
\label{warping_effects}
\end{figure*}

\section{Emergence of exceptional points at topological insulator-ferromagnet junction}

As we discussed, we consider a TI–FM junction. Importantly, for the TI, we consider a hexagonally-warped surface state. This is described by the following Hamiltonian proposed by Fu~\cite{fu2009hexagonal},

\begin{equation}
\begin{aligned}
H = \alpha(-k_y\sigma_x + k_x\sigma_y) + \lambda (k_x^3 - 3k_x k_y^2)\sigma_z .
\end{aligned}
\label{eq_TI}
\end{equation}

The first term represents the linear-in-momentum term with a Fermi velocity $\alpha$, with $k_x-k_y$ being the in-plane momenta. The second term corresponds to the hexagonal warping with strength $\lambda$. The warping term is invariant under three-fold rotations and plays an essential role to accurately describe the Fermi surfaces of rhombohedral bismuth based topological insulators~\cite{fu2009hexagonal,basak2011spin}. A comparison of the circular ($\lambda=0$) and hexagonal warped ($\lambda=1$) fermi surface is shown in Fig.~\ref{Fermi_surface}. At higher energies the warping becomes more pronounced and leads to the characteristic snowflake shape.

Including the self-energy of the FM lead, the complete junction can be represented by the following effective NH Hamiltonian,

\begin{equation}
 \begin{aligned}
        \Tilde{H} &= H + \Sigma_L\\
              &= \alpha(-k_y\sigma_x+ k_x\sigma_y)
              + \lambda(k_x^3-3k_xk_y^2)\sigma_z \\ 
              &+ \Sigma_L.   
\end{aligned}
\label{eq_TI_NH}
\end{equation}

Here, $\Sigma_L$ is the self energy term given by Eq.~\ref{Eq_self_energy}. This NH Hamiltonian can be expressed in the form $\Tilde{H} = \epsilon_0 + \mathbf{d} \cdot \boldsymbol{\sigma}$, where $\epsilon_0 = - i\Gamma$, $\epsilon_0 \in \mathbb{C}$ and the complex vector $\mathbf{d} = \mathbf{d}_R + i\mathbf{d}_I$, with $\mathbf{d}_R, \mathbf{d}_I \in \mathbb{R}^3$. For our specific case, the real part is given by $\mathbf{d}_R = (-\alpha k_y  \alpha k_x,  \lambda (k_x^3-3k_xk_y^2))$, and the imaginary part is $\mathbf{d}_I = (0,\ 0,\ -\gamma)$. The complex energy eigenvalues can be expressed in the form, $E_{\pm} = \epsilon_0 \pm \sqrt{\mathbf{d}_R^2 - \mathbf{d}_I^2 + 2i \mathbf{d}_R \cdot \mathbf{d}_I}$. NH degeneracies occur when the conditions $\mathbf{d}_R^2 = \mathbf{d}_I^2$ and $\mathbf{d}_R \cdot \mathbf{d}_I = 0$ are satisfied simultaneously. For the TI-FM junction with hexagonal warping, the degeneracy conditions read, 

\begin{equation}
\gamma^2 = \left(-\alpha k_y)^2\right) +(\alpha k_x)^2+ \lambda^2(k_x^3-3k_xk_y^2)^2,
\label{eq_EP_rules_a}
\end{equation}

\begin{equation}
\gamma\lambda(k_x^3-3k_xk_y^2) = 0.
\label{eq_EP_rules_b}
\end{equation}

Solving these equations, we find that NH degeneracies emerge in the TI–FM junction at the following locations in the $k_x$–$k_y$ plane,

\[
\textcolor{black}{
\left(0, \tfrac{\pm \gamma}{\alpha}\right)
\quad \text{and} \quad
\left(\tfrac{\pm \sqrt{3}\,\gamma}{2\alpha},\, \tfrac{\pm \gamma}{2\alpha}\right)
}
\]

In general, a total of six EPs appear. 
The hexagonal warping term plays a crucial role in formation of the exceptional degeneracies through the eigenvectors, as we will see. For the rest of our analysis, unless specified otherwise, we set \( \lambda = 1 \), \( \alpha = 1 \), \( \Gamma = 0 \), and \( \gamma = 1 \). To confirm that the degeneracies are indeed EPs, we plot the real and imaginary parts of the eigenvalues in Fig.~\ref{evals}(a) and (b), respectively, for zero magnetic field, $B_x=0$. The red dots mark the coalescence of the eigenvalues. Both real and imaginary parts of the energy eigenvalues coalesce at these points, thereby confirming the presence of EPs.

To further establish the coalescence of eigenvectors at these points, we evaluate the phase rigidity, $r$, as~\cite{heiss2012physics}

\begin{equation}
r = \frac{\langle \Psi_L | \Psi_R \rangle}{\langle \Psi_R | \Psi_R \rangle},
\label{phase_rigidity}
\end{equation}

where \( \Psi_L \) and \( \Psi_R \) denote the left and right eigenvectors, respectively. For an NH Hamiltonian, the Hilbert space is spanned by both left and right eigenstates~\cite{brody2013biorthogonal}, which are generally distinct, in contrast to the Hermitian case. Their orthogonality is established via a biorthogonal normalization, defined by \( \langle \Psi_L^{m} | \Psi_R^{n} \rangle = \delta_{mn} \). At an EP, the left and right eigenstates become mutually orthogonal, signaling the simultaneous coalescence of both eigenvalues and eigenvectors. Consequently, the phase rigidity vanishes near an exceptional degeneracy (\( r \to 0 \)) and approaches unity away from it. \textcolor{black}{Now, $\mathbf{d}$ can be expressed as}
\textcolor{black}{\[
\mathbf{d} = (d_x, d_y, d_z)
= \bigl(-\alpha k_y,\, \alpha k_x,\, \lambda (k_x^3 - 3 k_x k_y^2) - i\gamma \bigr).
\]}

\textcolor{black}{The corresponding right eigenfunctions are
\begin{equation}
\lvert \psi_\pm \rangle
=
\frac{1}{\sqrt{2E_\pm (E_\pm - d_z)}}
\begin{pmatrix}
d_x - i d_y \\
E_\pm - d_z
\end{pmatrix}.
\end{equation}}

\textcolor{black}{The only difference between the eigenstates is $E_+$ and $E_-$, and the energy eigenvalues are given by
\begin{equation}
E_{\pm}
=
\epsilon_0
\pm
\sqrt{
\alpha^2 (k_x^2 + k_y^2)
+
\bigl[\lambda (k_x^3 - 3 k_x k_y^2) - i \gamma \bigr]^2
}.
\end{equation}}
\textcolor{black}{
From these expressions, it follows that, both in the presence and absence of the warping term, the EP condition is characterized by $E_{+}=E_{-}$. At this point, the two eigenstates coalesce and become parallel, signaling the occurrence of an EP.} \textcolor{black}{We further plot the phase rigidity $r$ along the $k_x = 0$ line in Fig.~2(c) and find that it vanishes exactly at the EP locations $(0,\pm 1)$. We further confirm this by computing the vorticity around the EP, as shown in Fig.~2(d). \textcolor{black}{The analytical calculation of the vorticity yields $\pm \frac{1}{2}$, confirming the presence of second-order EPs (see Appendix A)}. The winding of the eigenvalues in the complex energy plane provides additional confirmation of the existence of the EPs.}
In Fig.~\ref{Field_effects}(a), we plot the phase rigidity, $r$, across the $k_x-k_y$ plane. We note that it forms a hexagonal pattern with $r\to 0$ at six discrete points. These are exactly our analytically obtained positions of the EPs (marked by the red dots), indicating the coalescence of eigenstates at these points. This hexagonal symmetry of the EPs is a direct consequence of the hexagonal warping term. \textcolor{black}{Next, we investigate the role of the warping term on the emergent EPs.}\\

\section{Shaping the non-Hermitian degeneracies with the warping term}
We begin with \( \lambda = 0 \), corresponding to the absence of the warping term. In this case, an exceptional ring is formed, where both the eigenvalues and eigenvectors coalesce along a closed ring, as shown in Fig.~\ref{warping_effects}(a). This behavior is analogous to the scenario discussed in Ref.~\cite{bergholtz2021exceptional}, where the addition of an imaginary term proportional to \( \sigma_z \) in a Dirac Hamiltonian leads to the formation of an exceptional ring. We now vary the warping strength and monitor the phase rigidity in the $k_x-k_y$ plane, as shown in Fig.~\ref{warping_effects}(b)–(d), corresponding to \(\lambda= 0.1\), \(0.5\), and \(1\), respectively. We find that, for all finite values of \(\lambda\), the exceptional ring fragments into six EPs. As the warping strength increases, the EPs remain settled at the vertices of a hexagon, reflecting the underlying symmetry imposed by the warping term. As such, the EPs are robust to the strength of the hexagonal warping. We note that the phase rigidity around the EPs is modulated by the warping strength, with the contours in Fig.~\ref{warping_effects}(b)-(d) becoming sharper as $\lambda$ is increased. \textcolor{black}{
In the absence of warping, i.e., for $\lambda = 0$, the Hamiltonian obeys chiral symmetry, $H = - \sigma_z^{\dagger} H \sigma_z$, provided there is no constant energy shift. Here, $\sigma_z$ acts as the chiral symmetry operator. This symmetry permits the emergence of exceptional rings, as reported in Ref.~\cite{yoshida2019symmetry,cayao2023exceptional}. For $\lambda \neq 0$, the chiral symmetry is broken, and consequently, the exceptional ring fragments into six isolated EPs.}\\

\begin{figure*}[t]
\includegraphics[width=1.0\textwidth]{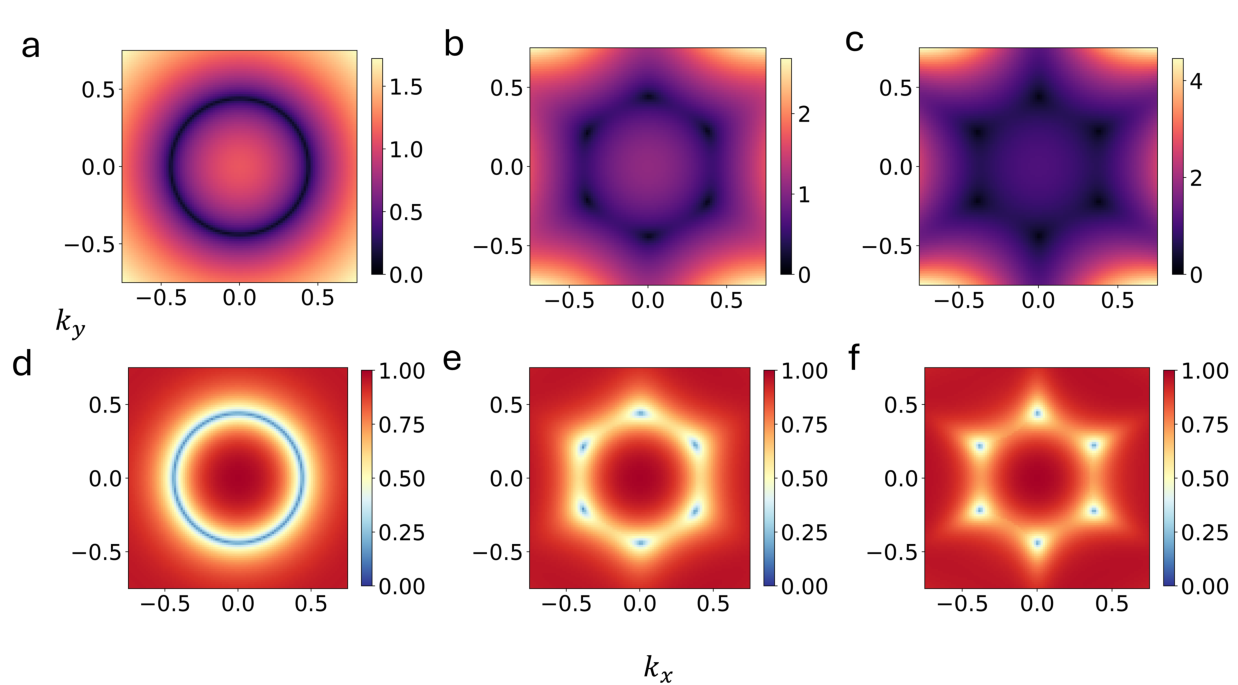}
\caption{\textbf{Exceptional ring fragmentation from lattice model.} The absolute energy difference between the two eigenvalues is shown in panels (a)–(c), while the corresponding phase rigidity is shown in panels (d)–(f), both plotted in the $k_x$-$k_y$ plane for $A_{12}=0$, $5$, and $10$, respectively. The bulk lattice model faithfully reproduces the behavior predicted by the low-energy theory. Specifically, the exceptional ring obtained in the absence of warping undergoes fragmentation into six EPs when finite hexagonal warping is introduced. The lattice model parameters are $A_{0}=0.0$, $B_{0}=0.0$, $A_{11}=2.0$, $B_{11}=1.0$, $B_{12}=3.0$, $A_{14}=0.30$, and $m_{11}=-10.0$, while the lead parameters are $t_{\parallel}^{L}=1.0$, $t_{z}^{L}=1.3$, $\mu_{\mathrm{L}}=4.0$, $m_{\mathrm{abs}}=0.5$, and $V_{\mathrm{SL}}=2.0$.}
\label{warping_effect_slab}
\end{figure*}

To complement our low-energy model results, we further investigate a microscopic lattice model. We consider the tight-binding model for Bi$_2$Se$_3$ class of topological insulators using the four-orbital tight-binding framework introduced in Ref.~\cite{mao2011tight}. In this lattice model, the five Dirac matrices that encode the spin-orbit and orbital couplings are defined as

\begin{equation}
\begin{aligned}
\Gamma_1 &= \sigma_1\otimes\tau_1, \qquad
\Gamma_2 = \sigma_2\otimes\tau_1, \qquad
\Gamma_3 = \sigma_3\otimes\tau_1,\\
\Gamma_4 &= 1\otimes\tau_2, \qquad
\Gamma_5 = 1\otimes\tau_3,
\end{aligned}
\label{eq:gamma}
\end{equation}

where $\sigma_i$ act on the spin and $\tau_i$ on the parity (orbital) degrees of freedom. The lattice is specified by the in-plane vectors
$a_1=(1,0)$, $a_2=(-\tfrac12,\tfrac{\sqrt3}{2})$, $a_3=(-\tfrac12,-\tfrac{\sqrt3}{2})$ and the out-of-plane (interlayer) vectors
$b_1=(0,\tfrac{\sqrt3}{3})$, $b_2=(-\tfrac12,-\tfrac{\sqrt3}{6})$, $b_3=(\tfrac12,-\tfrac{\sqrt3}{6})$, with the lattice constants set to unity ($a=c=1$). The bulk Bloch Hamiltonian takes the form

\begin{equation}
H_{\rm bulk}(\mathbf{k}) = h_0(\mathbf{k}) + \sum_{i=1}^{5} h_i(\mathbf{k})\,\Gamma_i,
\label{eq:bulk_H}
\end{equation}

where $\mathbf{k}=(k_x,k_y,k_z)$ is the three-dimensional crystal momentum. The coefficient functions are explicitly given by

\begin{align*}
h_0 &= 2A_0\,S(\mathbf{k}_\parallel) + 2B_0\,\Phi(\mathbf{k}_\parallel)\cos k_z, \\
h_1 &= -2A_{14}\sin\omega\,\bigl[\sin(\mathbf{k}\cdot a_2)-\sin(\mathbf{k}\cdot a_3)\bigr],\\
h_2 &= -2A_{14}\,\bigl[\sin(\mathbf{k}\cdot a_1)+\cos\omega\,(\sin(\mathbf{k}\cdot a_2)+\sin(\mathbf{k}\cdot a_3))\bigr], \\
h_3 &= 2A_{12}\sum_{i=1}^3 \sin(\mathbf{k}\cdot a_i), \\
h_4 &= -2B_{12}\,\Phi(\mathbf{k}_\parallel)\sin k_z, \\
h_5 &= 2A_{11}\,S(\mathbf{k}_\parallel) + 2B_{11}\,\Phi(\mathbf{k}_\parallel)\cos k_z + m_{11},
\end{align*}

with $\omega=-2\pi/3$, $\mathbf{k}_\parallel=(k_x,k_y)$, and the structure factors defined as $S(\mathbf{k}_\parallel)=\sum_{i=1}^3\cos(\mathbf{k}_\parallel\cdot a_i)$ and $\Phi(\mathbf{k}_\parallel)=\sum_{i=1}^3e^{i\mathbf{k}_\parallel\cdot b_i}$. The parameter $m_{11}$ controls the band inversion and thus determines the topological class, while the hopping amplitudes $A_{ij}$ and $B_{ij}$ describe the intra-layer and inter-layer couplings, respectively. Similar to the warping parameter $\lambda$ in the low-energy model, the parameter $A_{12}$ controls the strength of the hexagonal warping in this lattice model.\\

We consider a slab comprising $N=25$ quintuple layers stacked along the $z$ direction and attach a semi-infinite FM lead of identical geometry to one of its surfaces.

The self-energy (in spin space) is given by~\cite{bergholtz2019non},

\begin{equation}
\Sigma_{\mathrm{lead}}(\mathbf{k}_{\parallel})
=
\frac{V_{SL}^{2}}{t_{z}^{L}}
\left[
\frac{\sigma(\kappa_{+})+\sigma(\kappa_{-})}{2}\sigma_{0}
+
\frac{\sigma(\kappa_{+})-\sigma(\kappa_{-})}{2}
(\hat{\mathbf{m}}\cdot\boldsymbol{\sigma})
\right]
\otimes\tau_{0}.
\end{equation}

Where $S(\mathbf{k}_{\parallel})$ is the in-plane structure factor, with

\begin{equation}
\sigma(\kappa)=
\begin{cases}
\kappa-i\sqrt{1-\kappa^{2}}, & |\kappa|\le1,\\[4pt]
\kappa-\operatorname{sgn}(\kappa)\sqrt{\kappa^{2}-1}, & |\kappa|>1,
\end{cases}
\end{equation}

and,

\begin{equation}
\kappa_{\pm}(\mathbf{k}_{\parallel})
=
\frac{\mu_{L}-2t_{\parallel}^{L}S(\mathbf{k}_{\parallel})\pm m_{\mathrm{abs}}}
{2t_{z}^{L}}.
\end{equation}

Here, $t_{\parallel}^{L}$ and $t_{z}^{L}$ denote the in-plane and out-of-plane nearest-neighbor hopping amplitudes of the ferromagnetic lead, respectively, $\mu_{L}$ is the lead chemical potential, $m_{\mathrm{abs}}$ is the magnitude of the Zeeman splitting, $\hat{\mathbf{m}}$ is the unit vector specifying the magnetization direction, and $V_{SL}$ is the hopping amplitude describing the coupling between the lead and the TI slab.

Fig.~\ref{warping_effect_slab} shows the absolute difference between the energy eigenvalues and the phase rigidity in the $k_x$-$k_y$ plane for different values of the warping parameter. In the absence of hexagonal warping ($A_{12}=0$), the bulk lattice model hosts an exceptional ring, as evidenced by the vanishing energy splitting and the corresponding suppression of the phase rigidity along the ring. As the warping strength is increased, the continuous exceptional ring progressively fragments into six isolated EPs. These results nicely reproduce the behavior predicted by our low-energy effective theory, thereby demonstrating that the fragmentation of the exceptional ring into six EPs is a robust feature of the microscopic lattice model. Overall, our numerical results are consistent with the analytical low-energy model analysis.

\begin{figure*}[t]
\centering
\includegraphics[width=0.80\textwidth]{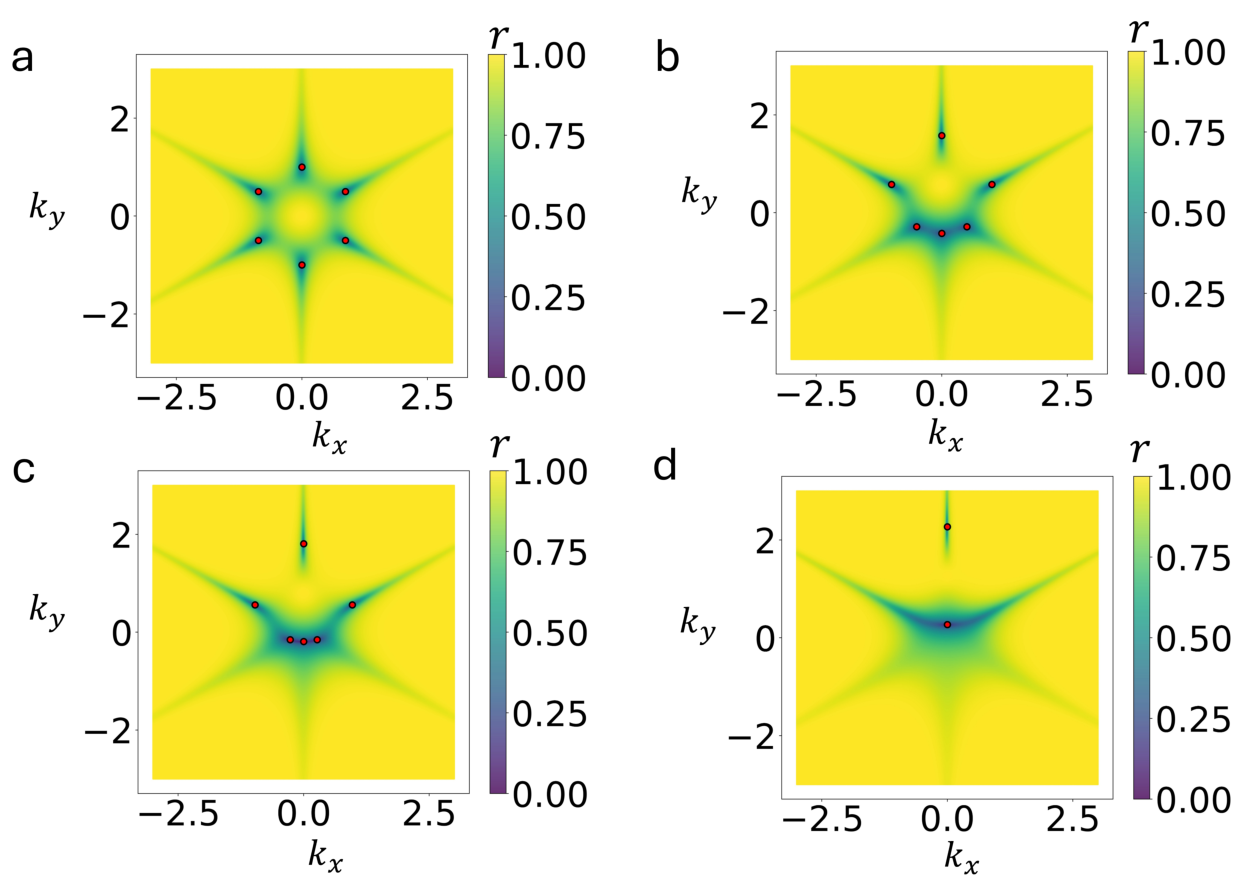}
\caption{\textbf{Magnetic-field tuning of exceptional points.} Evolution of the EP positions for different magnetic field values (a) $B_x=0.0B_c$, (b) $B_x=0.5B_c$, (c) $B_x=0.7B_c$, and (d) $B_x=1.1B_c$. The EPs are shown by red dots, while the phase rigidity is plotted in color (darker colors representing lower values of phase rigidity). As the magnetic field increases, the EPs move in momentum space, and above the critical field, $B_c$, four EPs annihilate, leaving behind two remanent EPs. The applied field also breaks the underlying hexagonal symmetry arising from the warping term. Here, we choose $\lambda=1$, $\alpha=1$,$\gamma=1$.}
\label{Field_effects}
\end{figure*}

\section{Tuning exceptional points with a magnetic field}
 
To study the magnetic-field control of the emergent EPs, we consider applying an in-plane magnetic field \( B_x \) along the interface, as illustrated in the schematic shown in Fig.~\ref{schematic}. \textcolor{black}{The corresponding Hamiltonian is given by,}
\begin{equation}
 \begin{aligned}
        H' &= \alpha(-k_y\sigma_x+ k_x\sigma_y)
              + \lambda(k_x^3-3k_xk_y^2)\sigma_z \\ 
              &+ \Sigma_L +B_x\sigma_x.   
\end{aligned}
\label{eq_TI_NH_B}
\end{equation}
\textcolor{black}{Where $B_x$ is an externally applied magnetic field along the $x$ direction (see Fig.~\ref{schematic})}
Notably, such a magnetic field allows delicate control over the positions of the emergent EPs. \textcolor{black}{The EP locations are given by,}

\[\textcolor{black}{
(0,\tfrac{B_x \pm \gamma}{\alpha}) 
\quad \text{and} \quad
\left(\pm \sqrt{3}\bar{k_y},\,
\bar{k_y}\right)},
\]
where, \textcolor{black}{$\bar{k_y}=\tfrac{B_x \pm \sqrt{4\gamma^2 - 3B_x^2}}{4\alpha}$.}

First, from our analytical expressions for the EP locations, we find that four of the EPs annihilate at the critical field \( B_c = \sqrt{\frac{4}{3}}\gamma \). This corresponds to the square root term in the location of the EPs vanishing. We monitor the evolution of the EPs with increasing magnetic field by means of the phase rigidity. The phase rigidity in the $k_x-k_y$ plane is plotted for a range of $B_x$ values in Fig.~\ref{Field_effects}(a)-(d). As we previously discussed, in the absence of a magnetic field there are six EPs. As the field strength is increased, we find that, among the six EPs, two move towards positive \( k_y \) direction on the \( k_x = 0 \) line, while the remaining four contract toward the origin.
\textcolor{black}{These four EPs form two distinct pairs. As $B_x$ increases, the EPs within each pair approach one another and eventually merge at $B_x = B_c$, where they coalesce. Exactly at the critical field, a total of four EPs exist.} For magnetic fields exceeding \( B_c \), only two EPs survive, which are located along the \( k_x = 0 \) line, as shown in Fig.~\ref{Field_effects}(d). We also note that a finite $B_x$ breaks the hexagonal symmetry of the EP positions. Applying the magnetic field along the \( y \)-direction leads to a qualitatively similar annihilation and motion of the EPs. Overall, we find that the interplay of the hexagonal warping and a magnetic field leads to a robust control over the EPs. \\

\section{Summary and outlook} 

We have studied the emergence of EPs at a TI–FM heterojunction, where the warping term plays crucial role in shaping the NH degeneracies. We show that the positions as well as the number of the arising EPs can be tuned by an applied magnetic field. Moreover, the EP locations inherit the hexagonal symmetry and are robust to the strength of the warping term of the TI surface state. \textcolor{black}{In the absence of the warping term, Bergholtz and Budich have suggested probing the surface conductance of the FM-TI junctions using a third normal lead or using surface spectroscopy techniques \cite{bergholtz2019non}. Similar setups may also be used in our proposal with hexagonal warping.} The prototypical topological material Bi$_2$Te$_3$, with its hexagonally warped surface states will be a suitable avenue for exploring our predictions~\cite{chen2009experimental,alpichshev2010stm}. Tuning the Fermi level by gating or doping could allow control over the degree of warping. Furthermore, heterostructures between Bi$_2$Te$_3$ and ferromagnets have also been recently fabricated~\cite{chen2022generation}. In summary, together with previous work, our study highlights TI–FM heterostructures, with hexagonally warped surface states, as a promising and tunable platform for realizing and probing NH phenomena. \\

\section*{Acknowledgments}
We thank A. Banerjee and A. Bose for useful discussions. M.A.R. is supported by a graduate fellowship of the Indian Institute of Science. A.N. acknowledges support from the DST MATRICS grant (MTR/2023/000021). \\

\section*{Author declarations}

\subsection*{Conflict of Interest}

The authors have no conflicts to disclose. \\

\subsection*{Author Contributions}

\noindent \textbf{Md Afsar Reja:} Investigation (lead); Methodology (lead); Software (lead); Writing - original draft (lead). \textbf{Awadhesh Narayan:} Conceptualization (lead); Methodology (supporting); Supervision (lead); Writing - review and editing (lead). \\

\section*{Data Availability}

The data that support the findings of this study are available from the corresponding authors upon reasonable request. \\

\renewcommand{\theequation}{A\arabic{equation}}
\renewcommand{\thesection}{A\arabic{section}}
\setcounter{equation}{0}
\setcounter{figure}{0}
\section*{Appendix A: Analytical calculation of vorticity}

We consider the Hamiltonian of a TI-FM junction with hexagonal warping and $B_x=0$,

\begin{equation}
H =
+ \alpha(-k_y \sigma_x + k_x \sigma_y)
+ \bigl[\lambda (k_x^3 - 3 k_x k_y^2) - i \gamma \bigr]\sigma_z .
\end{equation}

The eigenvalues are

\begin{equation}
E_{\pm} = - i \Gamma \pm D ,
\end{equation}

with $D = \alpha^2 (k_x^2 + k_y^2) + \bigl[\lambda (k_x^3 - 3 k_x k_y^2) - i \gamma \bigr]^2$. The energy difference $\Delta E = E_+ - E_- = 2D$ defines the vorticity

\begin{equation}
\nu = -\frac{1}{2\pi} \oint_{C} \nabla_{\mathbf{k}} \arg(\Delta E)\cdot d\mathbf{k}
= -\frac{1}{4\pi}\,\Delta \arg(D),
\end{equation}

where $C$ encircles an EP. We expand $D$ around the EP $(0,\gamma/\alpha)$. Let $k_x=u$ and $k_y=\gamma/\alpha+v$, with $u,v\ll1$. To linear order,

\begin{align}
\lambda (k_x^3 - 3 k_x k_y^2) &\simeq -3\lambda \frac{\gamma^2}{\alpha^2} u, \\
\alpha^2 (k_x^2 + k_y^2) &\simeq \gamma^2 + 2\alpha\gamma v .
\end{align}

Substituting into $D$ and keeping linear terms, we obtain,

\begin{equation}
D \simeq 2\alpha\gamma v
+ 6 i \gamma \lambda \frac{\gamma^2}{\alpha^2} u
= 2\alpha\gamma \bigl(v + i A u\bigr),
\end{equation}

where $A= \frac{3\lambda\gamma^2}{\alpha^3}$. Since $2\alpha\gamma>0$, the phase of $D$ is governed by $v+iAu$. On a small circular loop $u=r\cos\theta$, $v=r\sin\theta$,

\begin{equation}
v+iAu = r(\sin\theta + iA\cos\theta).
\end{equation}

The total change of its argument over $\theta\in[0,2\pi)$ is

\begin{equation}
\Delta \arg(D) = 2\pi\,\mathrm{sgn}(A).
\end{equation}

The vorticity is therefore

\begin{equation}
\nu = -\frac{1}{4\pi}\Delta \arg(D)
= -\frac{1}{2}\,\mathrm{sgn}(A).
\end{equation}

Since $\gamma^2/\alpha^3>0$, $\mathrm{sgn}(A)=\mathrm{sgn}(\lambda)$, yielding

\begin{equation}
\nu = -\frac{1}{2}\,\mathrm{sgn}(\lambda).
\end{equation}

The vorticity associated with the EPs are $\pm \tfrac{1}{2}$, reflecting the characteristic half-integer winding of EPs. \\

\renewcommand{\theequation}{D\arabic{equation}}
\renewcommand{\thesection}{D\arabic{section}}
\setcounter{equation}{0}
\setcounter{figure}{0}

\bibliography{ref.bib}
\vspace{0.5cm}

\end{document}